\documentclass[superscriptaddress,floatfix,amsmath,amssymb,aps,prb,twocolumn]{revtex4-2}

\usepackage{bbold}
\usepackage{comment}
\usepackage{graphicx}
\usepackage{dcolumn}
\usepackage{bm}
\usepackage[colorlinks=true,linkcolor=blue,citecolor=blue,urlcolor=blue]{hyperref}
\newcommand{\Tr}{\mathop{\rm Tr}\nolimits}
\renewcommand{\bm}{\mathbf}
\usepackage[usenames,dvipsnames]{xcolor}

\begin{document}

\title{Mechanisms of the in-plane magnetic anisotropy in superconducting NbSe$_2$}

\author{Menashe Haim}
\affiliation{Racah Institute of Physics, Hebrew University of Jerusalem, Jerusalem 91904, Israel}

\author{Alex Levchenko}
\affiliation{Department of Physics, University of Wisconsin--Madison, Madison, Wisconsin 53706, USA}

\author{Maxim Khodas}
\affiliation{Racah Institute of Physics, Hebrew University of Jerusalem, Jerusalem 91904, Israel}

\begin{abstract}
    We present a unifying picture of the magnetic in-plane anisotropies of two-dimensional superconductors based on transition metal dichalcogenides. 
    The symmetry considerations are first applied to constrain the form of the conductivity tensor. We hence conclude that the two-fold periodicity of transport distinct from the planar Hall related contributions, requires a tensor perturbation.
    At the same time, the six-fold periodic variation of the critical field results from the Rashba spin-orbit coupling on a hexagonal lattice.
    We have considered the effect of a weak tensor perturbation on critical field, gap function and magneto-conductivity.
    The latter is studied using the time dependent Ginzburg-Landau phenomenology. 
    The common origin of the $\pi$-periodicity in transport and thermodynamics properties is identified. 
    The scheme constructed here is applied to describe the existing theoretical scenarios from a unified point of view.  
    This allows us to single out the differences and similarities between the suggested approaches.
\end{abstract}

\maketitle

\section{Introduction}
The superconductivity in few-layer superconducting transition metal dichalcogenides (TMDs) is at the focus of the research for the last few years \cite{Lu2015,Ugeda2015,Saito2016,Xi2016,Costanzo2016,Dvir2017,DelaBarrera2018,Sohn2018,Hamill2021,Cho2020}.
The experimental work has been initially motivated by the progress in fabrication techniques resulting in the ability to exfoliate one to several stacked atomic layers on a substrate.
Surprisingly, these systems turned out to be superconducting with the critical temperatures of the same order of magnitude as in the bulk counterparts. 

\begin{figure}
\centering
\includegraphics[scale=0.35]{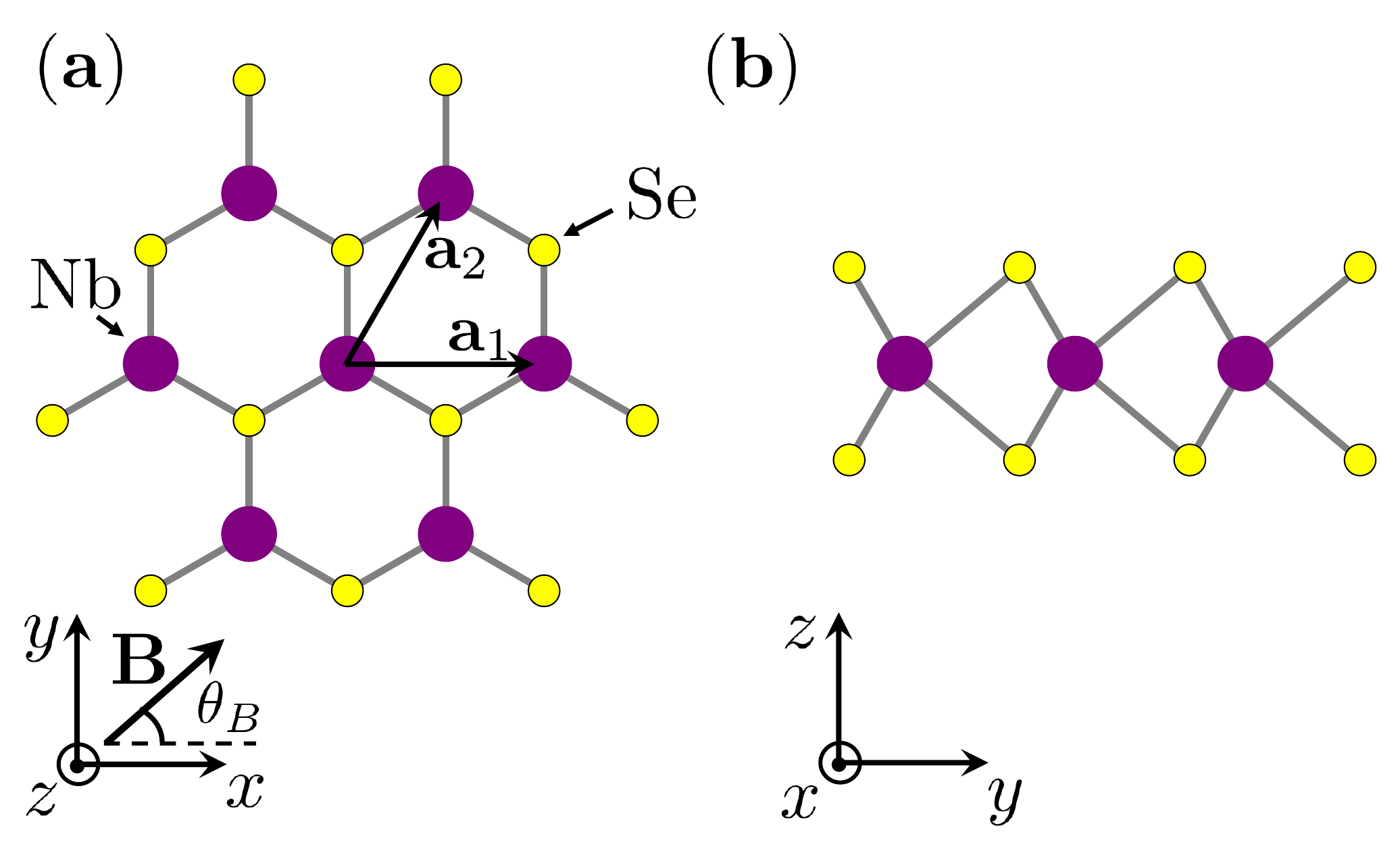}
\caption{\label{fig15} 
Crystal structure of $\mathrm{NbSe_{2}}$ mono-layer with bigger (purple) and smaller (yellow) circles denoting Nb and Se ions, respectively. 
Panel (a): top view of the crystal. $\mathbf{a}_{1},\mathbf{a}_{2}$ are primitive vectors of the triangular Bravais lattice. 
$x,y,z$ are Cartesian axes. 
The external field $\bm{B}$ lies in the $xy$-plane and forms the angle $\theta_B$ with the $x$ axis.
Panel (b): side view of the crystal.
}
\label{Fig-NbSe2}
\end{figure}

Samples with the odd number of layers including mono-layers such as exfoliated NbSe$_2$ and gated MoS$_2$ lack an inversion center, see Fig.~\ref{Fig-NbSe2}. 
This has two major consequences. 
First, the strong atomic spin-orbit (SO) coupling due to a transition metal splits electronic bands with the spin splitting larger than the superconducting gap by a few order of magnitudes \cite{Smidman2017}.

The strong SO interaction manifests itself in the strongly enhanced in-plane critical field, $B_c$ far above the usual Pauli limit \cite{Lu2015,Saito2016,Xi2016,Dvir2017,DelaBarrera2018,Sohn2018}. 
Thanks to the horizontal, in-plane mirror symmetry, $\sigma_h$ the SO interaction polarizes electrons out-of-plane, and is referred to as Ising SO coupling.
The superconductivity is protected in the Ising superconductor because in the presence of strong Ising SO coupling the in-plane spin susceptibility remains close to the Pauli susceptibility of a normal state \cite{Wickramaratne2020}. 

The second consequence of the lack of the inversion center, is the coexistence of the triplet superconductivity with the conventional $s$-wave singlet pairing \cite{Gorkov2001,Sigrist1991,Yip2014a}. 
In the case of Ising SO the electrons forming the triplet states have anti-parallel spins. Such triplet order parameter (OP), however decouples from the leading singlet OP when the SO splitting is much smaller than the Fermi energy, $E_F$ \cite{Frigeri2004d}.

More importantly, spin triplet correlations and SO coupling make it possible to manipulate the symmetry of the wave-function of the Cooper pairs by an external symmetry breaking perturbations.
For instance, in the TMD monolayers with prismatic coordination (Fig.~\ref{Fig-NbSe2}) the parallel spin triplets are induced by the in-plane field \cite{Mockli2019,Mockli2020}.
Such a triplet order parameter, is shown to affect the current-phase relation of a Josephson junction with the exchange interaction due to the ferromagnetic contacts \cite{Tang2021a}.
The Josephson current in this case depends on the angle between the magnetizations in the two ferromagnets.
Experimentally, the triplet OP might be related to the unusual field dependence of the gap at very high magnetic fields \cite{Kuzmanovic2021}.

As another example, the transformation of the OP by the externally applied magnetic field has been recently observed in the heavy fermion compound CeRh$_2$As$_2$ \cite{Khim2021}.
In this case the singlet pair density wave OP is favored by the magnetic field \cite{Yoshida2014}. 
While the magnetic field is pair breaking for the regular singlet OP, unconventional OPs can better adjust to it.

The unconventional OPs are rare, and it is interesting to explore the possibility of inducing such OP by externally applied perturbations. 
The reduced symmetry of the OP presumably causes anisotropy in the properties of a superconductor.
For this reason, such an anisotropy may potentially indicate the unconventional symmetry of the OP.
Very recently, the anisotropy has been reported in transport measurements in a few- and mono-layer NbSe$_2$ in the presence of an in-plane magnetic field $\bm{B}$ as a function of the field orientation \cite{Hamill2021,Cho2020}.
Indeed, the suggested interpretations build on a two-component spin-triplet OP either induced by external perturbation(s) or spontaneously formed, respectively. 

The one set of measurements studies the magnetic anisotropy in few-layer NbSe$_2$ device sandwiched between the two magnetic electrodes \cite{Hamill2021}.
This experiment reports the dependence of magneto-resistance on 
the angle $\theta_B$ formed by the in-plane field $\bm{B}$ and a fixed direction, Fig.~\ref{Fig-NbSe2}.
In the same setup, the critical field and the superconducting gap inferred from the tunneling data are measured.
For all three observables the data is $\pi$-periodic in $\theta_B$. 
This is in contrast to the six-fold symmetry expected from the underlying hexagonal crystallographic structure.
The $\pi$-periodic magneto-resistance is mostly observed in the transition region noticeably broadened by fluctuations and centered at $B_c$.
The $\pi$-periodicity of the gap function persists in the superconducting phase. 

In another set of experiments, \cite{Cho2020} the critical field anisotropy is measured in NbSe$_2$ mono-layers on a substrate.
In this case, the onset of the superconductivity at the critical field that exhibits a six-fold, $\pi/3$-periodicity. 
As the field is lowered the fully developed superconducting state sets in at the field that exhibits a $\pi$-periodicity. 
In this case the data has been interpreted in terms of two superconducting transitions.
The lower critical field has been argued to mark the nematic phase transition breaking the $C_3$ rotational symmetry spontaneously.  

Recently, a scenario of $\pi$-periodicity based on a conventional pairing has been suggested \cite{Wickramaratne2021}.
This approach has been originally motivated by the transport measurements in the tunnel junctions with the tunnel barrier made of an easy-axis ferromagnet separating the two Ising superconductors \cite{Kang2021}.
The magnetic hysteresis is tied to the superconductivity with the onset slightly below the critical temperature, $T_c$. 
This has been explained in terms of a different pair breaking efficiency of the magnetic impurities pointing in- and out-of-plane, respectively \cite{Mockli2020a}. 
The anisotropy of the magnetic scatterers  stabilized by the extended defects translates into the $\pi$-periodic critical field.  

In this paper we construct the symmetry based phenomenology that is general enough to capture the field dependence of transport and thermodynamic properties.
The goal of such a description is to contrast different scenarios of field anisotropy as well as to clarify their commonalities.

We assume that the leading superconducting instability is toward a singlet $s$-wave symmetric OP, $\psi$.
This does not exclude other subdominant pairing channels.
Moreover, at some point in the present analysis we specifically address them. 
The second assumption is that the two-fold symmetry is caused by the tensor perturbation, $\hat{\varepsilon}$.
In fact, we show that the tensor perturbation is necessary for the $\pi$-periodic trace of the conductivity tensor.

The $\hat{\varepsilon}$ tensor is assumed to be symmetric and without loss of generality traceless.
It appears as strain in the scenario of \cite{Hamill2021}, as the scattering anisotropy off magnetic impurities in \cite{Wickramaratne2021}.
This OP can also form spontaneously at the nematic transition as in \cite{Cho2020}.
Depending on the particular scenario $\hat{\varepsilon}$ can have a different physical realization.
For shortness, we will refer to it as strain.
To avoid the confusion we will state the meaning of $\hat{\varepsilon}$ explicitly whenever appropriate.

The main ingredient of our approach, is the coupling between the strain and the in-plane field.
To the leading order in $\bm{B}$, a finite strain changes the free energy by an amount
$\Delta F  \propto \Tr[(\bm{B}\bm{B})\hat{\varepsilon}]$, where $(\bm{B}\bm{B})$ denotes the dyadic tensor, with components, $(\bm{B}\bm{B})_{\alpha \beta} = B_{\alpha} B_{\beta}$.
Such term modifies the temperature dependence of the critical field, $B_c(T)$, and makes it two-fold anisotropic.
We show that the same contribution to the free energy explains the anisotrpic transport and thermodynamic properties.
At the same time it places certain restrictions on the microscopic mechanisms underlying this anisotropy.

The paper is organized as follows.
The findings that are independent of any particular scenario, and based solely on symmetry considerations are summarized in Sec.~\ref{sec:Main_Res}.
They include the discussion of $\pi$-periodicity of transport and thermodynamic properties. 
In Sec.~\ref{sec:Microscopic} we formulate some of the existing microscopic mechanisms of the field anisotropy within the general phenomenology introduced in Sec.~\ref{sec:Main_Res}.
In the concluding section \ref{sec:conclude} we discuss our results in light of the existing theories of field and/or strain induced anisotropy. 


\section{Symmetry considerations and Main Results}
\label{sec:Main_Res}
To clarify the restrictions imposed by symmetry it useful to extend the $D_{3h}$ point symmetry group of a TMD mono-layer to a $D_{\infty h}$ symmetry group of a finite circular cylinder.
In the systems with broken mirror symmetry the extended continuous symmetry we analyze is that of of the right cone, $C_{\infty v}$.
Having discussed the systems with the artificially extended continuous symmetries, we separately address the results depending on the discreteness of the symmetry group.  

\subsection{General form of the conductivity tensor}
\label{sec:sigma_gen}

Consider a two-dimensional system in the presence of the in-plane field.
As stated in the introduction we subject the system to tensor perturbation, $\hat{\varepsilon}_{\alpha\beta}$, $\alpha,\beta=x,y$. 
We stress again, that although we refer to $\hat{\varepsilon}$ as a strain for brevity, the only important assumption here is that it is a symmetric and traceless tensor.

The most general form of the in-plane conductivity tensor consistent with the Onsager relations, to linear order in strain and to all orders in $B$ is expressed as a sum of three contributions,
\begin{align}\label{eq:sigma_tensor_1}
    \hat{\sigma} = &  \mathbb{1} \left\{ \sigma_d + \sigma_{B\varepsilon}\Tr \left[(\bm{B}\bm{B})\hat{\varepsilon}\right]\right\} + \sigma_{\varepsilon} \hat{\varepsilon}
    \notag \\
    & + \sigma_p \left[(\bm{B}\bm{B}) - \mathbb{1}\Tr (\bm{B}\bm{B})/2  \right]\, ,
\end{align}
where $\mathbb{1}_{\alpha\beta} = \delta_{\alpha\beta}$ is a unit tensor.
Equation \eqref{eq:sigma_tensor_1} follows from the general theorems on invariants of the rotation group \cite{Weyl1966}.
Alternatively, it can be obtained by method of invariants for construction of material tensors \cite{Lax1974}. 
All coefficients, $\sigma_{d}$, $\sigma_{B\varepsilon}$, $\sigma_\varepsilon$ and $\sigma_p$ are some functions of $B^2$.

The first term of Eq.~\eqref{eq:sigma_tensor_1} represents the diagonal part of the conductivity tensor modified by a combined action of the magnetic field and strain.
The second term is a modification of the conductivity tensor due to strain alone. 

The off-diagonal part of the the last term, $\propto \sigma_p$ in Eq.~\eqref{eq:sigma_tensor_1} describes the dissipative planar Hall effect \cite{Goldberg1954,Koch1955}.
We refer to this whole term as a planar Hall contribution for shortness.
This term itself adds a $\pi$-periodicity in $\theta_B$ to all components of the conductivity tensor.
Therefore, care is needed to separate this $\pi$-periodicity from the one of the $\Tr{\hat{\sigma}}$.
The latter appears to the first order in $\hat{\varepsilon}$, while the former exists also at $\hat{\varepsilon}=0$.
Here we predominantly focus on the $\pi$-periodicity of the $\Tr{\hat{\sigma}}$.
We describe how the planar Hall may arise from the field induced anisotropy in Appendix~\ref{app:planar}.

The second consequence of Eq.~\eqref{eq:sigma_tensor_1} is that in scenarios, where the $s$-wave symmetry of the dominant pairing channel is not broken spontaneously, the $\pi$-periodicity sets in due to the combined action of the strain and the field.
In Eq.~\eqref{eq:sigma_tensor_1} it is described by the term proportional to $\sigma_{B\varepsilon}$ which has a form fixed by symmetry alone regardless of the microscopic details.

In the case of a free-standing or on-substrate mono-layer with the $D_{3h}$ or $C_{3v}$ symmetries respectively, the diagonal part of the conductivity has an angular dependence only at the sixth order in the field.
In both instances of the hexagonal symmetry, this is captured by the conductivity tensor written up to sixth order in the field and in the absence of other perturbations as follows,
\begin{align}\label{eq:sigma_D3h}
\hat{\sigma}  =& \mathbb{1} \left[\sigma_3 + \sigma_0 \mathrm{Re}( B_+^6 ) \right]
+
\sigma_2  \begin{bmatrix} \mathrm{Re}(B_+^2) & \mathrm{Im} (B_+^2) \\ \mathrm{Im} (B_+^2) & -\mathrm{Re}(B_+^2) \end{bmatrix} 
\notag \\
&+ 
\sigma_1
\begin{bmatrix} \mathrm{Re}(B_-^4) & \mathrm{Im} (B_-^4) \\ \mathrm{Im} (B_-^4) & -\mathrm{Re}(B_-^4) \end{bmatrix}\, ,  
\end{align}
where $\sigma_n(B^2)$ are polynomials of degree $n$ in $B^2$, $B_{\pm} = B_x \pm i B_y$, and we have set the $yz$-plane as the vertical mirror symmetry plane, Fig.~\ref{Fig-NbSe2}.

The terms proportional to $\sigma_3$ and $\sigma_2$ are the same in form as terms proportional to $\sigma_d$ and $\sigma_p$ in Eq.~\eqref{eq:sigma_tensor_1}, respectively.
The discreteness of the symmetry group allows the two additional terms in \eqref{eq:sigma_D3h}.
The term proportional to $\sigma_1$ is similar in structure to the planar Hall contribution present in both Eqs.~\eqref{eq:sigma_tensor_1} and \eqref{eq:sigma_D3h}, and has a $\pi/2$-periodicity.
We conclude from comparison of Eqs.~\eqref{eq:sigma_tensor_1} and \eqref{eq:sigma_D3h} that (i) the six-fold periodicity of $\hat{\sigma}$ requires the discrete hexagonal symmetry, and (ii) the two-fold periodicity in $\Tr \hat{\sigma}$ requires tensor-like perturbation.
These conclusions hold generally provided the $s$-wave symmetry of the order parameter is not broken spontaneously. 

\subsection{Thermodynamic properties}
\label{sec:Thermo}

We describe the thermodynamic properties via the Landau free energy functional, $F[\psi]$.
Here, $\psi(\bm{x})$ is the spin singlet, $s$-wave symmetry OP assumed to dominate other pairing channels. 
Keeping in mind the subsequent applications we allow for the spatial fluctuations of the OP described by the Fourier components, $\psi_{\bm{q}} = \mathcal{S}^{-1}\int d^2 \bm{x} e^{- i \bm{q}\bm{x}} \psi(\bm{x})$ with non-zero $\bm{q}$. 
Here, $\mathcal{S}$ stands for the area of a two-dimensional system.
For the present purposes we expand $F[\psi]$ up to fourth order in $\psi$,
$F[\psi] = \mathcal{S} \sum_{\bm{q}} E_{\bm{q}} |\psi_{\bm{q}}|^2 + c_4 \int d^2 \bm{x} |\psi(\bm{x})|^4$.
Here the fourth order coefficient, $c_4$ is taken as constant which only weakly depends on strain and the applied fields. 
Hence, we focus on the dispersion relation of the superconducting fluctuations, $E_{\bm{q}}$.

In the rotation invariant systems with a continuous symmetry group, $D_{\infty h}$ we have
\begin{align}\label{eq:E(q)}
\nu_0^{-1} E_{\bm{q}} = &  \epsilon + \beta B^2 + q^2 \xi^2   
+\!\left(\bm{B}\!\cdot\! \bm{q}\right)^2 \! \xi^2_B
+ \!\nu_0^{-1}\! E^{\varepsilon}_{\bm{q}}\, ,
\end{align}
where $\nu_0$ is the density of states per spin species,
$\epsilon = (T - T_c)/T_c$, $T_c$ is a zero field and zero strain critical temperature,
$\xi$ is a zero field coherence length, $\beta$ 
defines the zero strain critical field, $B_{c0} = \sqrt{-\epsilon/\beta}$.
In writing Eq.~\eqref{eq:E(q)} we normalize the OP, $\psi$ such that it coincides with the gap in the BCS limit.
The fourth term of Eq.~\eqref{eq:E(q)} describes the field induced anisotropy of the spectrum of fluctuations discussed in Appendix~\ref{app:planar}, and the last term is the correction to the spectrum of fluctuations due to strain, 
\begin{align}\label{eq:E_e(q)}
\nu_0^{-1} E^{\varepsilon}_{\bm{q}} & = 
\alpha_{B\varepsilon} \mathrm{Tr} [(\bm{B}\bm{B}) \hat{\varepsilon}]
+
\beta_{\varepsilon} \mathrm{Tr}[(\bm{q}\bm{q})\hat{\varepsilon}]
\notag \\
+ &
 \beta_{B\varepsilon } 
\mathrm{Tr} [(\bm{B}\bm{B}) \hat{\varepsilon}]
\left[
 B^2 q^2 \xi^2_{B \varepsilon }
+  \left(\bm{B}\!\cdot \!\bm{q}\right)^2  \xi'^2_{B \varepsilon }
\right]\, .
\end{align}
Here the first term yields the $\pi$-periodic modulation of the critical field, the second term describes the strain induced anisotropy of superconducting fluctuations.
The last term describes the $\pi$-periodic modulation of the dispersion.
As the strain perturbation is assumed to be weak, the terms of second order in momentum in the dispersion play a less significant role, and we omit them from the subsequent analysis. 

Eqs.~\eqref{eq:E(q)} and \eqref{eq:E_e(q)} hold in the $D_{\infty h}$ symmetric system with inversion symmetry.
In the system with $C_{\infty v}$ symmetry lacking the inversion center the dispersion relation also has  Lifshitz invariants linear in the momentum, $\bm{q}$ \cite{Smidman2017}.
These terms are crucial for a few effects predicted for the non-centrosymmetric superconductors such as onset of helical state, \cite{Mineev1994} and magnetoelectric effect, \cite{Edelstein1995,Yip2002}.
Here we assume that the $T_c$ enhancement due to the magnetoelectric effect is weak enough such that the field remains to be pair breaking.
Apart from the shift in the momentum the fluctuation spectrum is qualitatively the same for both continuous symmetries.
For this reason Lifshitz invariants present in $C_{\infty v}$ symmetric systems do not show up in the present calculation, and we omit them for clarity. 

The superconducting OP temperature and field dependence, $\psi(B,T)$ as well as the critical field temperature dependence, $B_c(T)$, easily follow from Eqs.~\eqref{eq:E(q)} and \eqref{eq:E_e(q)} evaluated at $\bm{q}=0$.
The OP reads,
\begin{align}\label{eq:psi}
\psi(T,B) = \sqrt{  -c_4^{-1}\left( \epsilon  +  \beta  B^2  +  \alpha_{B\varepsilon} \Tr [(\bm{B}\bm{B}) \hat{\varepsilon}] \right) }\, , 
\end{align}
where previously introduced $c_4$ coefficient is related to the zero field OP  
via $\psi(T<T_c,B=0) = \sqrt{- c_4^{-1}  \epsilon }$.
The OP given by Eq.~\eqref{eq:psi} is shown in Fig.~\ref{fig:res}(b) for $T<T_c$ and three different choices of the $\hat{\varepsilon}$ tensor.
Fig.~\ref{fig:res}(b) demonstrates that the amplitude and phase of the OP angular dependence is controlled by $\sqrt{\varepsilon_{xx}^2 +\varepsilon_{xy}^2}$, and $\varepsilon_{xx}/\varepsilon_{xy}$, respectively. 

Critical field is obtained from the condition of vanishing of the OP, $\psi(T<T_c,B_c)=0$.
This gives the relationship,
\begin{align}\label{eq:Bc}
\epsilon  +  \beta  B_c^2  +  \alpha_{B\varepsilon} \mathrm{Tr} [(\bm{B}_c\bm{B}_c) \hat{\varepsilon}]=0\, .
\end{align}
which can be easily solved 
\begin{align}\label{eq:Bc}
\frac{B_c(\theta_B)}{B_{c0}} = \left[ 1 - \alpha_{B\varepsilon} \frac{B_{c0}^2}{\epsilon}  
\left( \varepsilon_{xx} \cos 2 \theta_B + \varepsilon_{xy} \sin 2 \theta_B \right)
\right]^{-1/2}.
\end{align}

The critical field shown in Fig.~\ref{fig:res}(b) exhibits qualitatively similar angular dependence as the OP.
In both Eqs.~\eqref{eq:psi} and \eqref{eq:Bc} the term $\propto \alpha_{B\varepsilon}$ gives rise to the $\pi$-periodic oscillations with identical dependence of the oscillation phase on the strain orientation.

\subsection{Fluctuation mediated transport}
\label{sec:TDGL}

The $\pi$-periodic magneto-resistance has been reported at or near superconducting to normal transition driven by an in-plane field.
It is, therefore, rather plausible to relate this observation to the onset of critical fluctuations.

Here we present a phenomenological treatment of the transport at criticality based on the 
time dependent Ginzburg-Landau formulation.
Within this formulation we use the standard expression for the Aslamazov-Larkin fluctuation correction to the conductivity \cite{Aslamazov1968},
\begin{align}\label{eq:AL}
\sigma^{AL}_{\alpha\beta} =   e^2 \frac{\pi \nu_0}{4 T_c} 
    T  \int \frac{d^2 q}{(2 \pi)^2}  \frac{ v^{\alpha}_{\bm{q}} v^{\beta}_{\bm{q}} }{  E^3_{\bm{q}} }\, ,
\end{align}
where $v^{\alpha}_{\bm{q}} = \partial E_{\bm{q}}/\partial q_{\alpha}$ is the group velocity of the superconducting fluctuations, 
$e$ is electron charge, and here and below we set $k_B,\hbar =1$.
We have checked explicitly that the fluctuation correction captured by Eq.~\eqref{eq:AL} complies with the symmetry requirements expressed by Eq.~\eqref{eq:sigma_tensor_1} provided the spectrum of fluctuations takes the form given by Eqs.~\eqref{eq:E(q)} and \eqref{eq:E_e(q)}.

To see how the critical field anisotropy is related to the $\theta_B$ dependence of conductivity, it is instructive to consider the spectrum of fluctuations, 
\begin{align}\label{eq:E_appr}
   \nu_0^{-1} E_{\bm{q}} \approx   \epsilon + \beta B^2 + q^2 \xi^2   
+ \alpha_{B\varepsilon} \mathrm{Tr} [(\bm{B}\bm{B}) \hat{\varepsilon}]\, ,
\end{align}
which depends only on the magnitude of the momentum, $\bm{q}$, and yet is anisotrpic with respect to the field, thanks to the last term of Eq.~\eqref{eq:E_appr}. 
With the spectrum, \eqref{eq:E_appr}, the fluctuation correction, \eqref{eq:AL} 
can be written as $\sigma^{AL}_{\alpha\beta} = \delta_{\alpha\beta}  T e^2/16[T - T_c(B,\theta_B)]$, where the critical temperature renormalized by field and strain,
$T_c(B,\theta_B) = T_c - \beta B^2 - \alpha_{B\varepsilon} \mathrm{Tr} [(\bm{B}\bm{B}) \hat{\varepsilon}]$.

Experimentally, it is often more convenient to control the field $\bm{B}$ at fixed $T$.
For $T<T_c$ and $B > B_c$ we can rewrite the above result using the definition of $B_{c0}$ introduced earlier,  
\begin{align}\label{eq:AL1}
    \sigma^{AL}_{\alpha\beta}& = \delta_{\alpha\beta} \frac{e^2}{16} \frac{T}{T_c}
    \notag \\ 
 \times & \left[\left( 1- \frac{T}{T_c}\right)\left( \frac{B^2}{B_{c0}^2}- 1\right) + \alpha_{B\varepsilon} \mathrm{Tr} [(\bm{B}\bm{B}) \hat{\varepsilon}] \right]^{-1}\, .
\end{align}

\begin{figure}[h]
\centering
\includegraphics[scale=0.77]{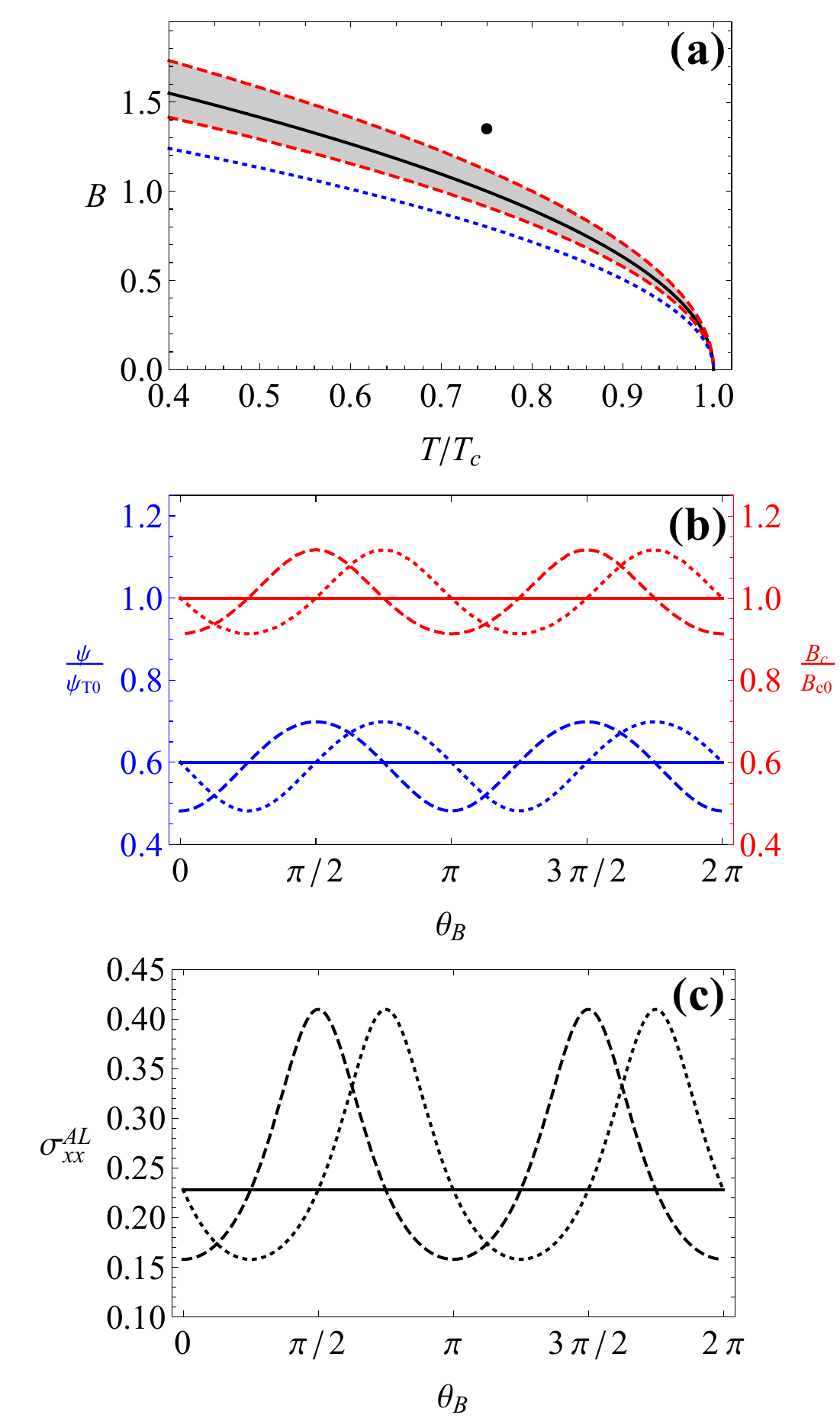}
\caption{
Panel (a): Solid (black) line shows the critical field without strain $B_{c0}$. 
Dashed (red) lines show the maximal and minimal critical field $B_{c}$, Eq.~\eqref{eq:Bc} attained as $\theta_B$ varies 
for
$|\alpha_{B\varepsilon}|\sqrt{\varepsilon^2_{xx}+\varepsilon^2_{xy} }/\beta=0.2$. 
The field is in units of $B_{c0}(\!T\!=\!0.75T_{c}\!)$.
Panel (b): Left axis (blue): the OP $\psi(\theta_B)$, Eq.~\eqref{eq:psi} for $B/B_{c0}=0.8$ [dotted (blue) line in panel (a)] normalized to $\psi_{T0} = \psi(B=0)$. 
Right axis (red):  $B_c(\theta_B)/B_{c0}$, Eq.~\eqref{eq:Bc}.
Here $\alpha_{B\varepsilon}\left(\varepsilon_{xx},\varepsilon_{xy}\right)/\beta=\left(0,0\right),\left(0.2,0\right),\left(0,0.2\right)$ for the solid, dashed and dotted lines, respectively.
All the curves are $T$ independent. 
Panel (c): $\sigma^{AL}_{xx}(\theta_B)/e^2$ per Eq.~\eqref{eq:AL1} for $B=1.35B_{c0}$ and $T=0.75T_{c}$ [black dot $\bullet$ in panel (a)] for 
$\alpha_{B\varepsilon}\left(\varepsilon_{xx},\varepsilon_{xy}\right)/\beta=\left(0,0\right)$,
$\left(0.2,0\right)$, 
$\left(0.2\right)$ 
shown by solid, dashed and dotted lines, respectively.
}
\label{fig:res}
\end{figure}

The conductivity correction given by Eq.~\eqref{eq:AL1} is shown in Fig.~\ref{fig:res}(c) for the set of parameters used in Fig.~\ref{fig:res}(b).
The result, \eqref{eq:AL1} suggests that the $\pi$-periodic $\theta_B$ dependence in the critical field, gap, and magnetoresistance have the same origin, expressed as a single term, $\propto \alpha_{B\varepsilon}$ in the free energy.
In fact, Figs.~\ref{fig:res}(b) and \ref{fig:res}(c) show that the maxima in $B_c$, $\psi$ and fluctuation conductivity  all occur at the same field orientation. 

The approximate spectrum of fluctuations, Eq.~\eqref{eq:E_appr} is isotropic with respect to momentum direction.
In result, the conductivity tensor, \eqref{eq:AL1} satisfies, $\sigma^{AL}_{xx} = \sigma^{AL}_{yy}$, and $\sigma^{AL}_{xy}=0$.
At finite field and/or strain the fluctuation spectrum is allowed to be anisotropic in momentum.
The microscopic origin of such an anisotropy requires a separate consideration that is beyond the scope of the current work.
Instead, in Appendix~\ref{app:planar} we show how the fluctuation spectrum anisotropy results in non-diagonal conductivity tensor in the form of the planar Hall effect. 


\section{Microscopic models of the critical field anisotropy}
\label{sec:Microscopic}
Here we discuss a potentially relevant microscopic mechanisms underlying the in-plane field anisotropy.

\subsection{Six-fold anisotropy}
Before addressing the two-fold anisotropy, for completeness we briefly discuss the possible origins of the pronounced six-fold anisotropy reported in Ref. \cite{Cho2020}.
It is rather natural that the SO interaction is necessary to couple Zeeman interaction to the six-fold anisotropy of the lattice.
Yet, the Ising SO coupling points out of plane with the in-plane field staying perpendicular to the spin polarization for all field orientations. 
We therefore do not expect the Ising SO taken alone to generate the angular dependence of the critical field.
This, of course is in agreement with the direct calculation.

In contrast, Ref. \cite{Kang2021} reports clearly different data for the in- and out-of-plane exchange fields. 
Similarly, for the purely in-plane field the anisotropy might result from the spin-polarization that can form different angles with the in-plane field. 
The well known SO coupling of this kind is the Rashba SO coupling.
It appears when the horizontal mirror symmetry, $\sigma_h$ is broken.
In fact, it has been shown in Ref.~\cite{Shaffer2020} that in the presence of the Rashba SO coupling the topological phase \cite{He2018}  is very sensitive to the direction of the magnetic field.
In particular, the two pairs of nodes present when the Zeeman splitting exceeds the superconducting gap, survive the Rashba SO only for the field aligned along the $\Gamma K$ directions.

Based on these observations we compute the angular dependence of $B_c$ in the presence of Ising and Rashba SO interaction within a minimal model of a single band superconductor with the band structure represented by the Hamiltonian, 
\begin{equation}
H_{0}=\underset{\bm{k},s}{\sum}\xi_{\bm{k}}c_{\mathbf{k}s}^{\dagger}c_{\bm{k}s}+\underset{\bm{k},ss'}{\sum}\left[\boldsymbol{\gamma}_{\bm{k}}-\bm{B}\right]\cdot\boldsymbol{\sigma}_{ss'}c_{\bm{k}s}^{\dagger}c_{\bm{k}s'},
\label{eq:H0}
\end{equation}
where $\xi_{\bm{k}}$ is the energy measured from $E_F$, $\boldsymbol{\gamma}_{\bm{k}}=-\boldsymbol{\gamma}_{-\bm{k}}$ is the SO coupling term, $c_{\mathbf{k}s}^{\dagger}$ creates a particle with the momentum, $\mathbf{k}$ and spin $s$.
We denote by $\boldsymbol{\sigma}=\left(\sigma_{1},\sigma_{2},\sigma_{3}\right)$
the vector of Pauli matrices. 

For a specified SO coupling, $\boldsymbol{\gamma}_{\bm{k}}$, the critical field, $B_c(T)$ is determined by the solution of the linearized self-consistency equation
\cite{Frigeri2004,Frigeri2004d,Frigeri2006}
\begin{align}\label{eq:self_gamma}
& \ln\left(\frac{T}{T_{\mathrm{c}}}\right)  +
\pi T\sum_{n=-\infty}^{\infty}  \Bigg[\frac{1}{\left|\omega_{n}\right|}
\notag \\
& -   \left\langle  \frac{\left|\omega_{n}\right|\left(\boldsymbol{\gamma}^{2}+\omega_{n}^{2}\right)}{\omega_{n}^{2}\left(B_c^{2}+\gamma^{2}\right)+(\bm{B}_c\!\cdot \! \boldsymbol{\gamma})^2+\omega_{n}^{4}}\right\rangle _{\mathrm{F}}\Bigg] = 0 ,
\end{align}
where $\langle \ldots \rangle_{\mathrm{F}}$ stands for the angular averaging over the Fermi surface,
$\omega_n = \pi T (2 n +1)$, $\bm{B}_c = B_c (\bm{B}/B)$.

\begin{figure}
\centering
\includegraphics[scale=0.84]{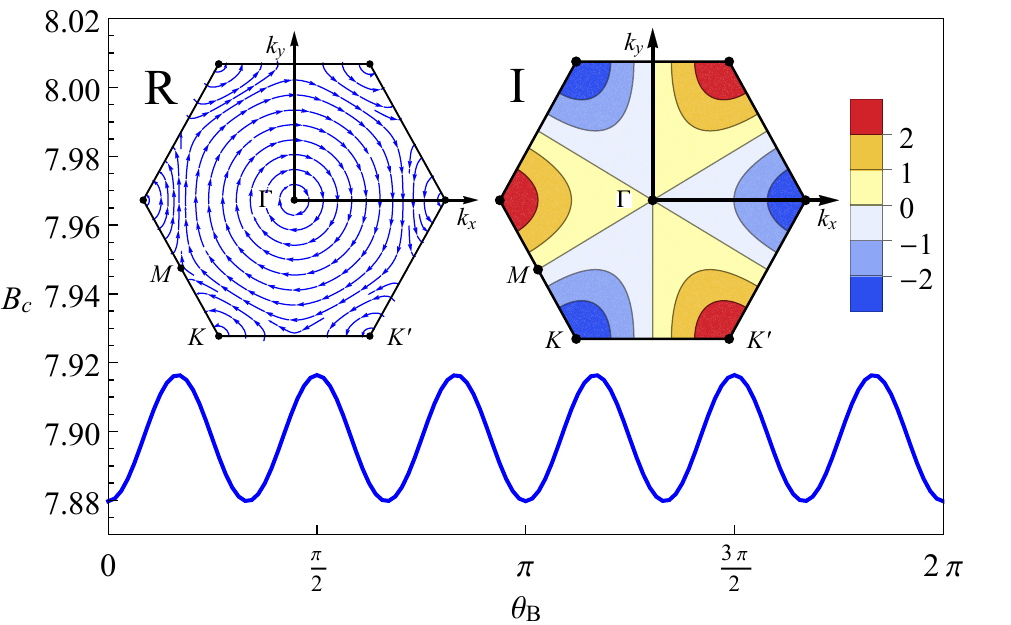}
\caption{\label{fig:6fold} 
The six-fold oscillations of the critical field, $B_c$ in units of $T_c$ as a function of the field direction, $\theta_B$ obtained by solving Eq.~\eqref{eq:self_gamma} with $T=0.8T_{c},\gamma^{I}=80T_{c},\gamma^{R}=4T_{c},k_{\mathrm{F}}a_{1}=2.9$.
Insets show the direction of the Rashba SO coupling (R) and the magnitude of the Ising SO coupling (I) aligned with the $z$ axis in units of $\gamma^{I}$ shown in the first Brillouin zone.
}
\end{figure}

We write, $\boldsymbol{\gamma}_{\bm{k}} = \boldsymbol{\gamma}^I_{\bm{k}} + \boldsymbol{\gamma}^R_{\bm{k}}$, 
with the dominant SO coupling of Ising type, $\boldsymbol{\gamma}^I_{\bm{k}}$ and much weaker Rashba SO interaction, $\boldsymbol{\gamma}^R_{\bm{k}}$.
For simplicity, we employ the tight-binding single band approximation,  
\begin{subequations}
\label{eq:gamma}
\begin{align}\label{eq:gamma_I}
    \boldsymbol{\gamma}^I_{\bm{k}} & = \gamma^I \hat{z} [\sin(\bm{k}\cdot \bm{d}_1) + \sin(\bm{k}\cdot \bm{d}_2) + \sin(\bm{k}\cdot \bm{d}_3)]
    \end{align}
    \begin{align}\label{eq:gamma_R}
    \boldsymbol{\gamma}^R_{\bm{k}} & = \gamma^R \frac{\sqrt{3}}{2} \hat{x} [ \sin(\bm{k} \cdot\bm{d}_2) - \sin(\bm{k} \cdot\bm{d}_3)] 
    \notag \\
    + \gamma^R & \frac{1}{2}\hat{y}[ \sin(\bm{k} \cdot\bm{d}_2) + \sin(\bm{k}\cdot \bm{d}_3) - 2 \sin(\bm{k}\cdot \bm{d}_1 )]\, , 
\end{align}
\end{subequations}
where the vectors, $\bm{d}_i$ are expressed via the Bravais lattice vectors, $\bm{a}_{1,2}$ shown in Fig.~\ref{Fig-NbSe2} as $\bm{d}_1 = \bm{a}_1$, $\bm{d}_2 = \bm{a}_2-\bm{a}_1$, and  $\bm{d}_3 = -\bm{a}_2$.
Eqs.~\eqref{eq:gamma} can be obtained based on the representations of $D_{3h}$ and $C_{3v}$ groups on the hexagonal lattice, respectively \cite{Smidman2017}. 
The critical field obtained from Eqs.~\eqref{eq:self_gamma}, \eqref{eq:gamma} are illustrated in Fig.~\ref{fig:6fold}.
From symmetry we expect the six fold modulation to appear in sixth order in field, which would make the effect rather small, as indeed is apparent from Fig.~\ref{fig:6fold} for the typical choice of parameters.

The origin of oscillations is deduced from the observation made in Ref.~\cite{Shaffer2020} regarding the stability of the topological phase.
The oscillations arise because the Ising SO coupling Eq. \eqref{eq:gamma_I} vanishes along $\Gamma M$.
Indeed, for such momenta the spin-splitting is solely due to Rashba SO Eq. \eqref{eq:gamma_R}.
The superconductivity $\bm{k}$ on $\Gamma M$ lines is least protected when $\boldsymbol{\gamma}^R_{\bm{k}} \parallel \bm{B}$. 
We stress that the present analysis does not relate directly to the field induced topological phase, since the above calculation is performed in the normal state.

\subsection{Mechanisms of two-fold anisotropy}

The natural question is whether the existing physically motivated models conform to the phenomenology presented thus far.
We, indeed find this to be correct.
In some cases this is not automatic, and we furnish some restrictions on these models. We start with the discussion of the models formulated in terms of the conventional OP.

\subsubsection{Anisotropic magnetic impurities}
\label{sec:MI}

In the recent work Ref. \cite{Wickramaratne2021} the scenario of two-fold field anisotropy has been suggested based on the effect of magnetic defects with an easy axis.
This point of view has an added advantage of explaining the hysteretic behaviour tied to the superconductivity \cite{Kang2021} rather naturally.
Moreover it lends itself to the phenomenology presented above.

Indeed, the isotropic part of the pair breaking effect of the field captured by the constant $\beta$ in Eq.~\eqref{eq:E(q)} \cite{Mockli2020a},
\begin{align}\label{eq:beta}
    \beta = & \pi T\sum_{n=-\infty}^{\infty}
  \frac{\Gamma_{m}'\left(\Gamma_{m}'^{2}+(\gamma^I)^{2}\right)}
  {\left(2\Gamma_{m}+\left|\omega_{n}\right|\right)^{2}\left(\left|\omega_{n}\right|\Gamma_{m}'+(\gamma^I)^{2}\right)}
     \notag \\
& \times      
    \frac{1}{\left[\Gamma_{m}'\left(2\Gamma_{m}+\left|\omega_{n}\right|\right)+(\gamma^I)^{2}\right]}\, ,
\end{align}
where $\Gamma_m$ is the scattering rate off the magnetic impurities and $\Gamma_{m}' = \Gamma_m + |\omega_n|$. Here and in what follows, we considered the Ising SO coupling, Eq.~\eqref{eq:gamma_I} for the momenta close to the $K$-points of the Brillouin Zone.
In this case the SO coupling takes the form, 
$\boldsymbol{\gamma}\left(\varphi_{\mathbf{k}}\right) \approx \hat{z} \gamma^I   \mathrm{sgn\left[\cos\left(3\varphi_{\mathbf{k}}\right)\right]}$.
Although this approximation applies to $K$-pockets, the results for the $\Gamma$ pocket are qualitatively similar.

What is crucial for us here is that in addition to $\beta$ we have $\alpha_{B\varepsilon} \hat{\varepsilon} \neq 0$ signifying the field anisotropy when the magnetic impurities have an easy axis. 
The coefficient of the first term of Eq.~\eqref{eq:E_e(q)} controlling the two-fold anisotropy reads,  
\begin{align}\label{eq:alpha}
    \alpha_{B\varepsilon} \hat{\varepsilon} =  &
     \pi T \Bigg( \sum_{n=-\infty}^{\infty} \frac{-\Gamma_{m}\Gamma_{m}'^{2}}{\left(2\Gamma_{m}+\left|\omega_{n}\right|\right)^{2}\left[\left|\omega_{n}\right|\Gamma_{m}'+(\gamma^I)^{2}\right]} 
    \notag \\
& \times      
    \frac{1}{\left[\Gamma_{m}'\left(2\Gamma_{m}+\left|\omega_{n}\right|\right)+(\gamma^I)^{2}\right]}
    \Bigg)\hat{\varepsilon}_{\varphi}
\end{align}
where, 
\begin{align}\label{eq:e_phi}
    \hat{\varepsilon}_{\varphi}=\begin{bmatrix}
    \cos 2 \varphi & \sin 2 \varphi \\ \sin 2 \varphi & -\cos 2 \varphi 
    \end{bmatrix}\, 
\end{align}
contains the dependence of the critical field on the easy axis direction, specified by the angle $\varphi$ it forms with $x$-axis.
Equation \eqref{eq:e_phi} reflects the transformation property of a second rank tensor, as the easy axis direction changes.
Put simply, it ensures that the two-fold anisotropy enters via the combination $\propto \cos[2(\theta_B - \varphi)]$ which is naturally expected. 
We emphasize that the physically meaningful quantity is the product, $\alpha_{B\varepsilon} \hat{\varepsilon}$.
In the presented scenario $\hat{\varepsilon}$ does not have a meaning of strain. 
It is therefore, neither possible nor necessary to consider $\alpha_{B\varepsilon}$ and $\hat{\varepsilon}$ separately in this case.

\subsubsection{Coupling between the leading singlet and subleading triplet channels}
\label{sec:ST}
Following Ref. \cite{Hamill2021} we now consider the possibility of the two-fold anisotropy arising from the coupling between the leading $s$-wave instability and the subleading unconventional triplet OP(s).
Our approach here remains the same. 
It, again builds upon observation that the relevant observable is the critical field.

Consider a two-component triplet order parameter, $\bm{\eta} = (\eta_1,\eta_2)$.
The free energy including the two OPs can be written as 
\begin{align}\label{eq:coupl}
    F[\psi,\bm{\eta}] = & \epsilon |\psi|^2 + \epsilon_t (|\eta_1|^2 + |\eta_2|^2) +c_4 |\psi|^4
    \notag \\
    & + 
    \left[ \psi^* \sum_{l=1}^2  C^*_{l}(\bm{B},\hat{\varepsilon}) \eta_l  + c.c. \right]\, ,
\end{align}
where $\epsilon_t = ( T - T_{t})/T_t$, $T_t < T_c$ is the critical temperature of the triplet channel.  
The free energy in Eq.~\eqref{eq:coupl} is minimized with respect to $\eta_{i}$ for
$\eta_i = - \psi C_i(\bm{B},\hat{\varepsilon}) / \epsilon_t $.
Substitution of this solution to Eq.~\eqref{eq:coupl} gives the effective free energy,
\begin{align}\label{eq:coupl1}
    F[\psi] = & \left[ \epsilon  - \epsilon^{-1}_t \sum_l | C_l |^2 \right] |\psi|^2 +c_4 |\psi|^4
   \, 
\end{align}
describing the condensation of the singlet OP.

The question at this junction is how the free energy in Eq.~\eqref{eq:coupl1} may result in $\pi$-periodic critical field. 
It can appear via the specific dependence of the coupling coefficients $C_l$ on the field and strain.
The gap function is a mixture of the isotropic singlet and anisotropic triplet components.
Still, in the considered scenario the thermodynamic state retains the symmetry of the underlying lattice.
The same is true for the free energy in Eq.\eqref{eq:coupl1}.
Therefore, the $\pi$-periodicity follows if the coefficient $ \sum_l | C_l |^2$ happens to generate the combination, $\propto \Tr[(\bm{B}\bm{B})\hat{\varepsilon}]$.
For this to happen, the couplings $C_l$ should contain two kinds of terms,
$C_l = C^B_l + C^{\varepsilon B}_l$, where $C^B_l \propto B$, and yet additionally,
$C^{\varepsilon B}_l$ linear in both $B$ and $\hat{\varepsilon}$.

To be specific, we consider a $D_{3h}$ symmetric system where the $A_1'$ symmetric singlet coexists with $E''$ field induced parallel spin triplets \cite{Mockli2019}.
To the linear order in the field the coupling constants in this case  are fixed by the symmetry, $C^B_l \propto i(\hat{z}\times \bm{B})_l$.  
To write the symmetry allowed coupling linear in both strain and the field,
note that the vector, $\hat{\varepsilon}\bm{B}$ with components, $(\hat{\varepsilon}\bm{B})_l = \sum_{l'} \hat{\varepsilon}_{ll'}B_{l'}$ belongs to $E''$ as does the $\bm{B}$.
Hence, $C_l = \lambda_1 i(\hat{z}\times \bm{B})_l + \lambda_2 i(\hat{z}\times \hat{\varepsilon}\bm{B})_l$, where $\lambda_{1,2}$ are two constants. 
Consulting Eq.~\eqref{eq:coupl1} we obtain in the considered scenario, 
\begin{align}
    \alpha_{B\varepsilon} = - 2 \epsilon_t^{-1} \lambda_1 \lambda_2\, .
\end{align}

The question arises as to the microscopic origin of the coupling proportional to both $\hat{\varepsilon}$ and $B$.
One possible assumption leading to such a coupling is that the strain renormalizes the Zeeman interaction such that it becomes
\begin{align}\label{eq:H_zr}
    H_{\mathrm{eff}} = \bm{B}_{\mathrm{eff}}\cdot \boldsymbol{\sigma}\, ,\,\,\, \bm{B}_{\mathrm{eff}} =\bm{B} + \lambda \hat{\varepsilon}\bm{B}\, .
\end{align}
The form of Eq.~\eqref{eq:H_zr} is fixed by symmetry.
This, however is insufficient to estimate the relative importance of the $g$-factor anisotropy.
For that reason we describe the possible mechanism of the correction to the $g$-factor due to the tensor perturbation in some details in Appendix~\ref{app:atomic}.
We have estimated 
$ 
\lambda \approx (\lambda_{\text{SO}}/\Delta E_{cr})
    \left( \Delta E_{str}/\Delta E_{cr} \right)
$,
where $\lambda_{\text{SO}}$ is the atomic SO coupling strength, $\Delta E_{cr}$ is a crystal field splitting, and $\Delta E_{str}$ is a typical energy scale associated with the strain.

It follows that $\lambda$ in Eq.~\eqref{eq:H_zr} is sensitive to the microscopic details such as, for instance, the splitting between the $e_g$ and $t_{2g}$ orbitals even and odd under the mirror, $\sigma_h$, respectively.
Generally, the ratio $ \Delta E_{str}/\Delta E_{cr}$ is expected to be small.
However, it might not be small near the extended defects, which may lead to modification of the OP  or even the local time reversal symmetry breaking \cite{Willa2021}.

\begin{figure}
\centering
\includegraphics[scale=0.48]{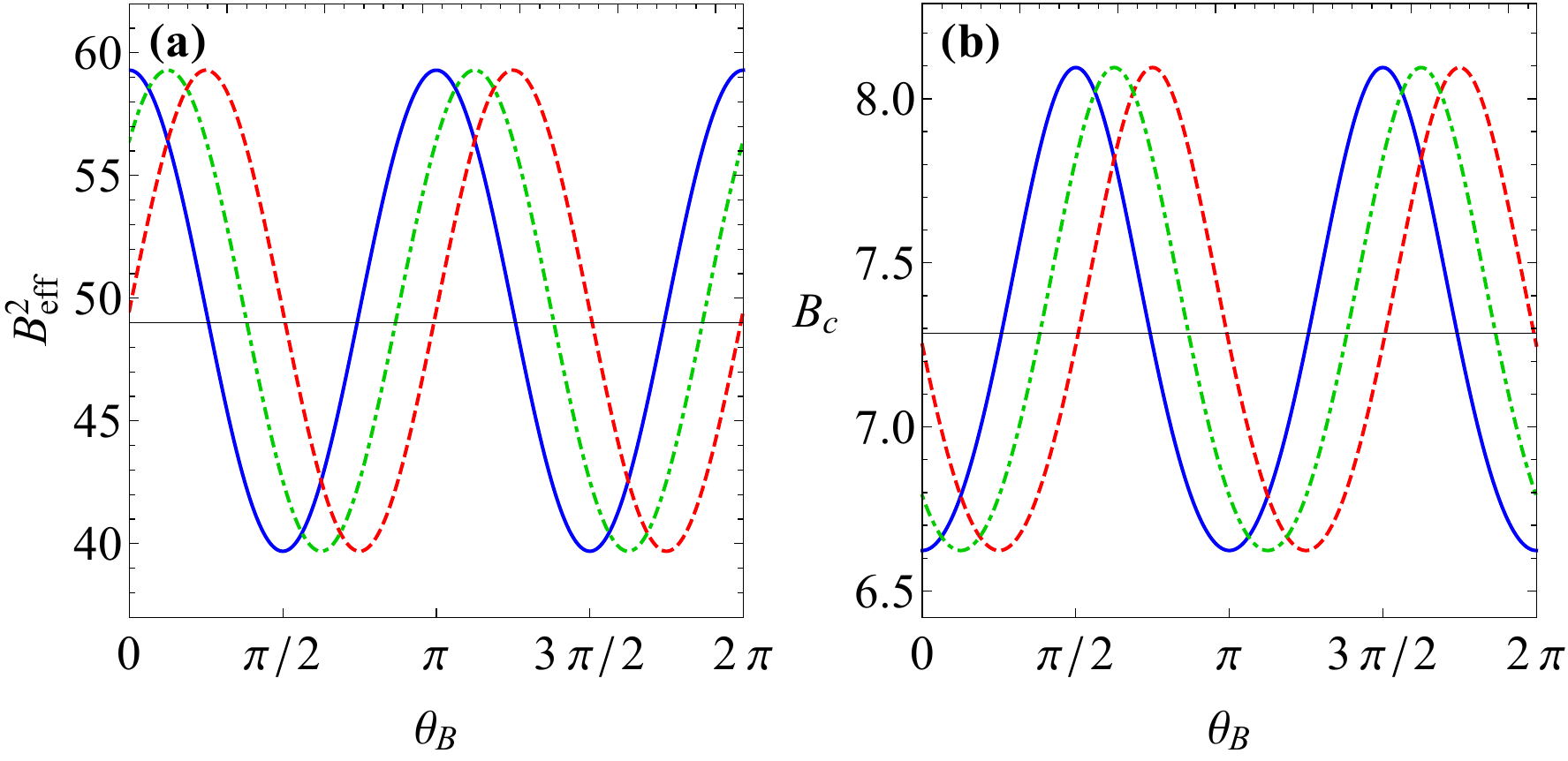}
\caption{\label{fig:Beff} Panel (a): The angular dependence of the effective magnetic field $B_{\mathrm{eff}}^{2}$, computed from the definition Eq.~\eqref{eq:H_zr}. The magnitude of the magnetic field $\left|\mathbf{B}\right|=7T_{c}$. Panel (b): The angular dependence of the critical field $B_c$ obtained by solving Eq.~\eqref{eq:self-cons_Bc} for the clean case with $T=0.5T_{c}$, $\gamma^I= 15 T_{c}$.
Panels (a,b):
The results obtained for the values of $\lambda\left(\varepsilon_{xx},\varepsilon_{xy}\right)=$ 
$\left(0,0\right)$, $\left(0.1,0\right)$, $\left(0,0.1\right)$ and $\left(0.1,0.1\right)/\sqrt{2}$ are shown by the thin (black), solid (blue), dashed (red) and dotted-dashed (green) lines, respectively.
All curves are $\pi$-periodic and have a phase difference of $\pi/8$. 
$B_c$ and $B_{\mathrm{eff}}$ are given in units of $T_c$.}
\end{figure}

One consequence of Eq.~\eqref{eq:H_zr} is that the two-fold anisotropy results already in the scenario based on a single component, $s$-wave OP.
Indeed, the pair breaking effect is caused by the Zeeman splitting of the electronic bands.
With the effective Zeeman interaction, Eq.~\eqref{eq:H_zr} the spin splitting itself becomes anisotropic.
This is illustrated in the Fig.~\ref{fig:Beff}a for different strain tensor, $\hat{\varepsilon}$. 
We, therefore, address this possibility in the next section.

\begin{figure}
\centering
\includegraphics[scale=0.45]{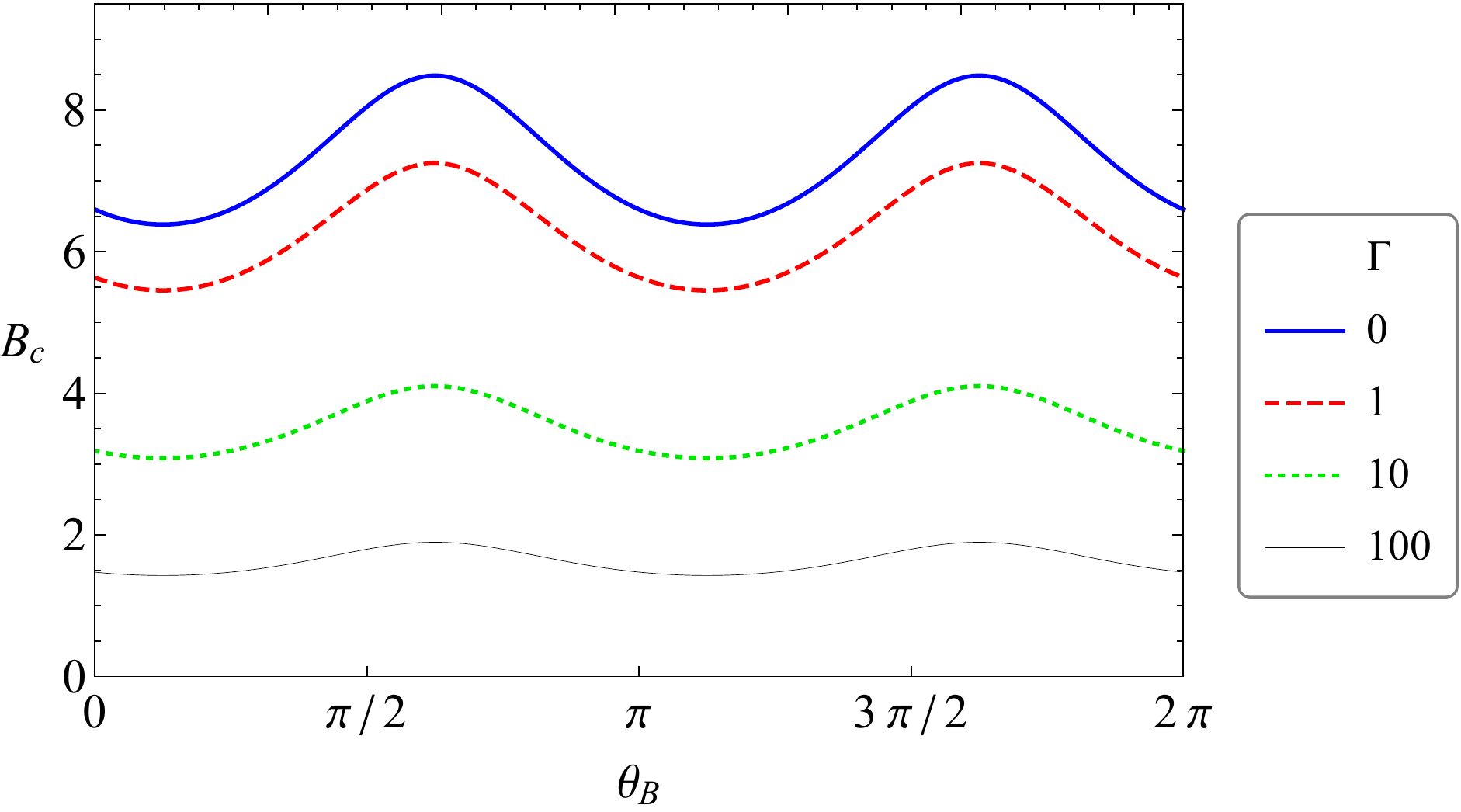}
\caption{\label{fig:dirty} The angular dependence of the critical field for different values of impurity scattering rate $\Gamma$, obtained from Eq.~\eqref{eq:self-cons_Bc} for 
$T=0.5T_{c}$, $\gamma^I=15T_{c}$, $\lambda\left(\varepsilon_{xx},\varepsilon_{xy}\right)=\left(0.1,0.1\right)$. The results obtained for $\Gamma=0,1,10,100T_{c}$ are shown in solid (blue), dashed (red), dotted (green) and thin (black) lines respectively. As in the clean case, Fig.~\ref{fig:Beff}b, $B_{c}\left(\theta_{B}\right)$ has a $\pi$ periodicity. 
The angular dependence is suppressed as the disorder scattering rate, $\Gamma$ increases. $B_c$ is given in units of $T_c$.}
\end{figure}

\subsubsection{Critical field anisotropy induced by the anisotropic $g$-factor} 
\label{sec:induced}

Now we make an assumption of an anisotropic $g$-factor given by Eq.~\eqref{eq:H_zr}, and study how this generates the $\pi$-periodic critical field $B_c$. 
We adopt the same strategy as before to describe the angular dependence of the critical field.
Specifically, we work within the phenomenological scheme presented in the Sec.~\ref{sec:Main_Res}.
In the expression for the free energy given by Eq.~\eqref{eq:E(q)} we have for the constant controlling the critical field, $B_{c0}$,
\begin{align}\label{eq:beta_anis}
    \beta = \pi T\sum_{n=-\infty}^{\infty}  
    \frac{\left(\Gamma+\left|\omega_{n}\right|\right)}{\omega_{n}^{2}\left[(\gamma^I)^{2}+\Gamma\left|\omega_{n}\right|+\omega_{n}^{2}\right]}, 
\end{align}
and because of the Eq.~\eqref{eq:H_zr}, we have a simple relationship,
\begin{align}\label{eq:alpha_anis}
    \alpha_{B\varepsilon} = 2 \lambda \beta\, .
\end{align}

In fact, for the particular scenario of the field anisotropy based on Eq. \eqref{eq:H_zr} we can compute the critical field without making an expansion in $B$ (see Appendix~\ref{app:Bc} for details),
\begin{align}\label{eq:self-cons_Bc}
& \ln\left(\frac{T}{T_{c}}\right) +\pi T\sum_{n=-\infty}^{\infty}
\bigg\{
\frac{1}{\left|\omega_{n}\right|}
\notag \\
& - \frac{(\gamma^I)^{2}+\Gamma\left|\omega_{n}\right|+\omega_{n}^{2}}
{\left|\omega_{n}\right|\left[B_{\mathrm{eff}}^{2}+(\gamma^I)^{2}+\omega_{n}^{2}\right]+\Gamma\left(B_{\mathrm{eff}}^{2}+\omega_{n}^{2}\right)}
\bigg\}
 =0,
\end{align}
where we have allowed for non-magnetic disorder characterized by the elastic scattering rate $\Gamma$.

We solve Eq.~\eqref{eq:self-cons_Bc} numerically for different choices of the anisotropic part of the $g$-factor tensor, $\lambda \hat{\varepsilon}$.
The results are shown in Fig.~\ref{fig:Beff}b.
In the present approach the anisotropy of the critical field is a direct consequence of the anisotropy of the $g$-factor Eq. \eqref{eq:H_zr}.
This is illustrated by the juxtaposition of the angular dependence of the effective field, $B_{\mathrm{eff}}$ Eq. \eqref{eq:H_zr} for an external field, of a fixed magnitude and the angular dependence of the critical field.
Comparison of Fig.~\ref{fig:Beff}a and \ref{fig:Beff}b shows that maximal (minimal) $B_c$ occurs for minimal (maximal) $g$-factor.

In the present scenario, the scalar disorder randomizing different directions of motion tends to suppress the effect of the $g$-factor anisotropy.
The detrimental effect of the disorder scattering on the critical field anisotropy is illustrated in Fig.~\ref{fig:dirty}.

\subsubsection{Two-fold periodicity resulting from the nematic transition}

In this scenario, suggested in Ref. \cite{Cho2020} the role of tensor perturbation $\hat{\varepsilon}$ is played by the components $(\eta_1,\eta_2)$ of the triplet OP, that is assumed to form spontaneously.  
In this scenario, taking for instance the system with $C_{3v}$ or $D_{3h}$ symmetry, the contribution to the free energy that gives rise to $\pi$-periodicity reads, $\propto 2 B_x B_y \eta_1 + (B_x^2 - B_y^2) \eta_2$.
The rest of the analysis is then similar to the above, with a similar outcome.

\section{Conclusions}
\label{sec:conclude}

We have constructed a phenomenological theory of the in-plane magnetic field anisotropy in two-dimensional TMD based superconductors.
The starting point of the discussion is the analysis of constrains imposed on the conductivity tensor by symmetry to all orders in the magnetic field.
The symmetry alone implies that the two-fold anisotropy of the trace of the conductivity tensor requires a symmetry breaking tensor perturbation.
Alternatively, such a tensor perturbation may result if the superconductivity breaks the symmetry of the underlying lattice, e.g. via a nematic phase transition.

The individual entries of the conductivity tensor may have a two-fold anisotropy because of the standard planar Hall effect.
On experimental level, therefore, it is important to differentiate between the $\pi$-periodicity of the trace of the conductivity tensor and $\pi$-periodicity related to the planar Hall contribution.
The very same discussion makes it clear that the six-fold anisotropy requires either $D_{3h}$ or $C_{3v}$ symmetries.

We then turned to the thermodynamic properties, focusing initially on the single component $s$-wave OP.
We have identified the specific combination of the tensor perturbation and the magnetic field that is responsible for $\pi$-periodicity in both transport and thermodynamic properties. 
This has allowed us to formulate the existing scenarios of the $\pi$-periodicity within the same scheme.
Such a reformulation reveals the limitations of the existing approaches, their commonalities and differences. 

In the approach of Ref. \cite{Hamill2021} we have found the specific form of the coupling between the leading spin-singlet and subleading spin-triplet OPs required for $\pi$-periodicity.
We have consequently described a way such coupling can be realized.
In the theory of Ref. \cite{Wickramaratne2021} the tensor perturbation results from the anisotropy in the scattering properties of magnetic impurities stabilized by extended defects. 
In this case the tensor perturbation is not related at least directly to the strain.
In the scenario of Ref. \cite{Cho2020} the role of the tensor perturbation is played by the components of a triplet OP formed spontaneously. 

Clearly, more detailed studies are required to fully clarify the origin of in-plane field anisotropy in TMD based few-layer systems under different external conditions. We believe that the presented phenomenological theory may serve as a convenient framework in addressing the related questions.

\subsection*{Acknowledgments}

We thank M. Aprili, R. Fernandes, I. Mazin, K. Michaeli, D. M\"ockli , V. Pribiag, C. H. L. Quay, 
H. Steinberg and
D. Wickramaratne for useful discussions on various topics related to this study. 
We especially thank M. Smolkin for the illuminating discussions of the invariants allowed by the continuous symmetries.
M.H. and M.K. acknowledge the financial support from the Israel Science Foundation, Grant No. 2665/20.
The work at UW-Madison was financially supported by the U.S. Department of Energy (DOE), Office of Science, Basic Energy Sciences (BES) Program for Materials and Chemistry Research in Quantum Information Science under Award No. DE-SC0020313 (A.L.).

\begin{appendix}
\section{Anisotropy of fluctuation spectrum and planar Hall effect}
\label{app:planar}

Here we trace the relation between the anisotropy of the fluctuation spectrum and the finite planar Hall effect.
To this end, we compute the fluctuation conductivity from Eq.~\eqref{eq:AL} yet now with $ \xi_{B} \neq 0$ in the dispersion of the superconducting fluctuations, Eqs.~\eqref{eq:E(q)} and \eqref{eq:E_e(q)}.
This term describes the anisotropy of the dispersion relation of superconducting fluctuations due to the finite magnetic field.  
Such an anisotropy gives rise to the fluctuation induced planar Hall effect.
This contribution is contained in the general expression, Eq.~\eqref{eq:sigma_tensor_1} as a term proportional to $\sigma_p$.

\begin{figure}
\centering
\includegraphics[scale=0.55]{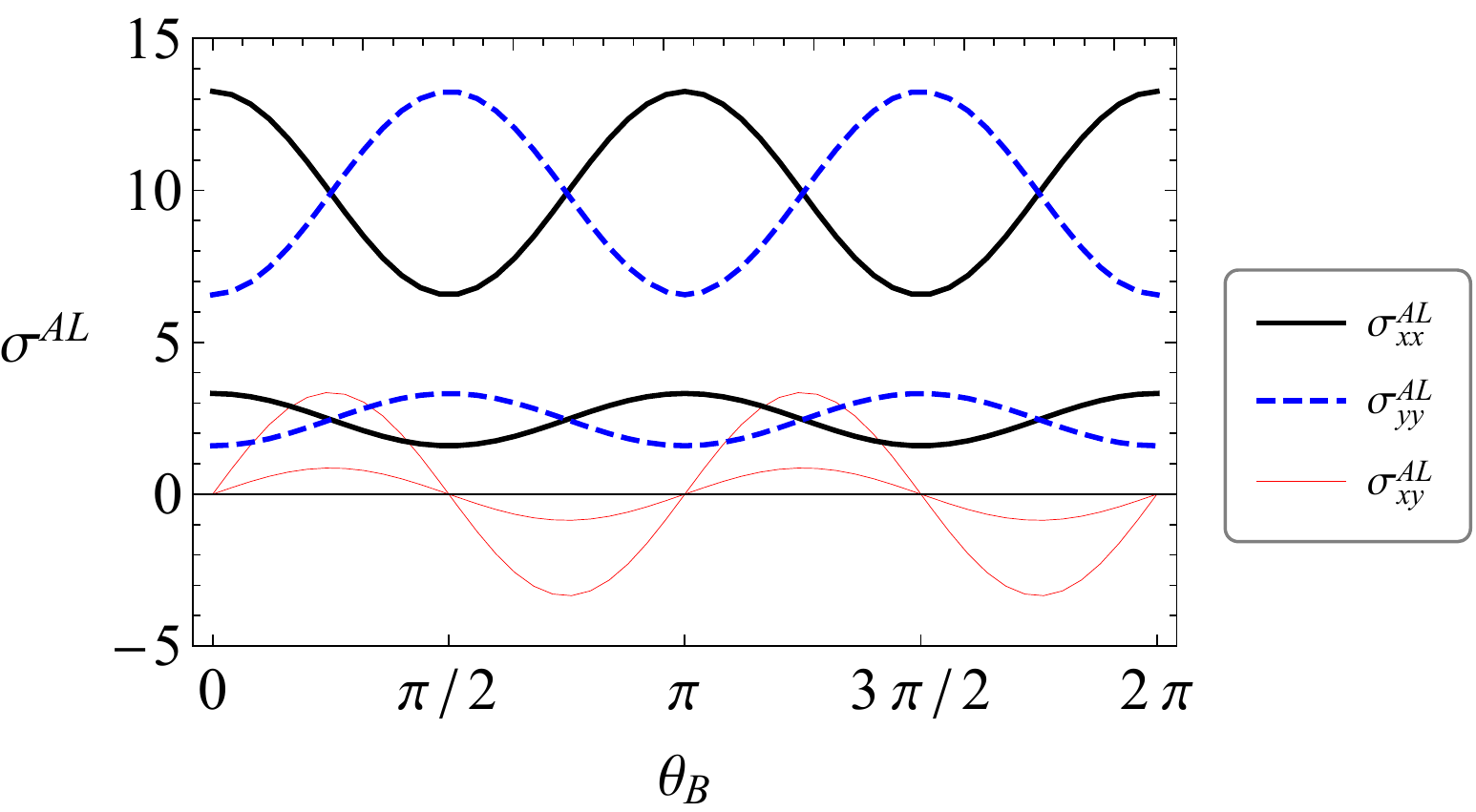}
\caption{\label{fig:sigma_pi} 
Fluctuation planar Hall effect due to the field induced fluctuation spectrum anisotropy. 
$\sigma^{AL}$ in units of $e^2$ is computed from Eq.~\eqref{eq:AL} for 
$T=0.75T_{c},\xi_{B}B_{c0}/\xi=1$, and $\alpha_{B\varepsilon}=\beta_{\varepsilon}=\beta_{B\varepsilon}=0$.
 The solid (black), dashed (blue) and thin (red) lines denote the $\sigma^{AL}_{xx}$, $\sigma^{AL}_{yy}$ and $\sigma^{AL}_{xy}$ components of $\sigma^{AL}$, respectively.
 Each of the three components is shown for $B=1.04 B_{c0}$ and $B =1.01 B_{c0}$. 
 The fluctuation correction, $\pi$-periodicity, and the planar Hall (anisotropic) part of the conductivity tensor become more pronounced as the field approaches the critical one.
 }
\end{figure}

The result is presented in Fig.~\ref{fig:sigma_pi}.
Clearly, the planar Hall contribution is $\pi$-periodic, and exhibits an enhancement for the field and/or the temperature approaching the transition.
We stress that such a contribution has to be disentangled from the $\pi$-periodicity of $\Tr \hat{\sigma}$.
The microscopic origin of the field induced spectrum anisotrpies is beyond the scope of the present work.

\section{Strain induced anisotrpy of the atomic $g$-factor}
\label{app:atomic}

 Here we argue that the strain like perturbation described by a tensor $\hat{\varepsilon}$ gives rise to the anisotropy of the Zeeman coupling expressed as Eq.~\eqref{eq:H_zr}. 
 Such a $g$-factor anisotropy most readily follows in the atomic limit.
For once an atom is subject to strain-like perturbation its response to a Zeeman field is expected to become anisotrpic in the presence of the SO coupling.

To illustrate the idea consider the electronic states at the $\Gamma$ point predominantly having a character of $d_{z^2}$ orbitals.
At $\Gamma$ the two states $|0,1/2\rangle$ and $|0,-1/2\rangle$ are spin degenerate.
The in-plane field $\bm{B}$ couples these states via the usual Zeeman interaction, 
$H_z = \bm{B}\cdot \boldsymbol{\sigma}$, where the Pauli matrices, $\bm{\sigma} = (\sigma_x,\sigma_y,\sigma_z)$ act in the subspace of the two states, $|0,\pm 1/2\rangle$.
We subject this system to the strain-like perturbation, $\mathcal{H}_{\varepsilon}$.
Here we are not interested in its exact form.
What matters is its transformation properties under the symmetry operations.
Hence, we write $\mathcal{H}_{\varepsilon} = \sum_{i,j} \hat{\varepsilon}_{ij} X_{i}X_{j}$, we denote by $X_i$ any vector operator.
Such a perturbation causes the virtual transitions to the states with different orbital content.
Our goal is to show that such transitions modify the effective Hamiltonian acting in the space $|0,\pm 1/2\rangle$ thus taking the form of Eq.~\eqref{eq:H_zr}.

\subsection{Atomic Hamiltonian and crystal field}
\label{app:atomic_H}
Consider the $d$-shell atomic levels of a transition metal ion.
For definiteness, we consider the limit of crystal field being stronger than the SO coupling.
Neglecting for a moment the SO coupling the crystal field lifts the five fold orbital degeneracy of a $d$-shell into the $A_{1g}$, $E_{1g}$ and $E_{2g}$.
These orbitals appropriate to the $D_{\infty h}$ symmetry are
characterized by the $z$-component of the angular momentum, $m=0$, $m=\pm 1$ and $m=\pm 2$, respectively.

The crystal field quenches the in-plane components of the angular momentum, $L_{x,y}$, while the expectation value of $L_z$ in the orbital states listed above staying finite.
For this reason we represent the atomic Hamiltonian in the form, 
$\mathcal{H}_a = \mathcal{H}_0 + V$, where the perturbation reads,
\begin{align}\label{eq:V}
    V =  \lambda_{\mathrm{SO}} (L_x \sigma_x + L_y \sigma_y)+g_{L}\bm{B}\cdot\bm{L} +\mathcal{H}_{\varepsilon}\, .
\end{align}
The unperturbed Hamiltonian, $\mathcal{H}_0$ describes the bare atomic $d$-shell atomic level structure, and includes the crystal field effects as well as a part of the SO interaction, $\lambda_{\mathrm{SO}} L_z \sigma_z$ left unquenched by the crystal field.
In addition, the last but one term in Eq.~\eqref{eq:V} contains a usual coupling of the magnetic field to the orbital motion of an electron in the atom.

Our approach here is to consider the terms other than $\mathcal{H}_0$ as a small perturbation.
This is justified in the limit of crystal field being stronger than SO coupling.
Indeed, in this case the $\lambda_{\mathrm{SO}} L_z \sigma_z$ as part of $\mathcal{H}_0$ lifts the double spin degeneracy.
The states then form five Kramers doublets $|0,\pm 1/2\rangle$, $|\pm 1, \pm 1/2 \rangle$, $|\pm 1, \mp 1/2 \rangle$, $|\pm 2, \pm 1/2 \rangle$, $|\pm 2, \mp 1/2 \rangle$, where the state $|m,s\rangle$, has the z-component of the spin, $s= \pm1/2$.
The doublets transform as $E_{1/2g}$, $E_{3/2g}$, $E_{1/2g}$, $E_{5/2g}$, and $E_{3/2g}$ spinor representation of the double group, $D_{\infty h}$.

\subsection{Perturbation theory and effective Hamiltonian}
\label{sec:PT}
For definiteness, we consider the space of $|0,\pm 1/2\rangle$ as appropriate to the electronic states residing at the hole pocket centered at $\Gamma$.
Equation \eqref{eq:H_zr} is an effective Hamiltonian describing spectrum in the above space of two states.

By adopting the results of Ref.~\cite{Shavitt1980} to the present problem we obtain for the matrix elements of the effective Hamiltonian, apart from the original Zeeman splitting,
\begin{align}\label{eq:3rd}
    H^{\text{eff}}_{ss'} & = \!\! \sum_{{m,m' \neq 0\atop s_{1,2}}}\!\!
   (E_{0s} - E_{ms_1})^{-1}(E_{0s} - E_{m's_2})^{-1}
   \notag \\
\times &    \langle 0s|V_X|m s_1\rangle \langle m s_1|V_D|m' s_2\rangle \langle m' s_2|V_X|m  s_1\rangle\, ,
\end{align}
where $E_{ms}$ are the unperturbed energies defined by $\mathcal{H}_0$.
In Eq.~\eqref{eq:3rd} the perturbation, \eqref{eq:V} is split into a diagonal and off-diagonal parts, $V = V_X + V_D$ defined in terms of the projection operator, 
$\mathcal{P} = |0,1/2 \rangle \langle 0,1/2| + |0,-1/2 \rangle \langle 0,-1/2|$ as
$V_X = \mathcal{P}\mathcal{H} (1 - \mathcal{P}) + (1 - \mathcal{P})\mathcal{H}\mathcal{P}$, 
$V_D = V-V_X$.

It is convenient to rewrite the perturbation Hamiltonian, \eqref{eq:V} in the form
\begin{align}\label{eq:V1}
    V =& \lambda_{\mathrm{SO}} (L_+ \sigma_- + L_- \sigma_+ ) + g_L (B_+ L_- + B_- L_+) 
    \notag \\
        &    + (\varepsilon_- \bar{L}_+^2 + \varepsilon_+ \bar{L}_-^2)
\end{align}
where, $\sigma_{\pm} = \sigma_x \pm i \sigma_y$, $L_{\pm} = (L_x \pm i L_y)/2$,
$\bar{L}_{\pm} = X_1 \pm i X_2$ transforming as $L_{\pm}$, and 
$\varepsilon_{\pm} = \varepsilon_{xx} - \varepsilon_{yy} \pm 2 i \varepsilon_{xy}$.
With Eq.~\eqref{eq:V1} the effective Hamiltonian takes the form,
\begin{align}
   & H^{\text{eff}} = 
    \sigma_+ 
     \frac{\langle 0|\lambda_{\mathrm{SO}} L_-|1\rangle \langle 1 |g_L L_-|2 \rangle \langle 2 |\varepsilon_{-}\bar{L}^2_{+}|0 \rangle }{(E_{0,1/2} - E_{1,-1/2})(E_{0,1/2} - E_{2,1/2})}
     \\
      + &
     \sigma_+ 
     \frac{\langle 0|g_L L_-|1\rangle \langle 1 |\lambda_{\mathrm{SO}} L_-|2 \rangle \langle 2 |\varepsilon_{-}\bar{L}^2_{+}|0 \rangle }{(E_{0,1/2} - E_{1,1/2})(E_{0,1/2} - E_{2,-1/2})}+\ldots
     + h.c.,\notag
\end{align}
where $|m\rangle$ is the orbital state with the out-of-plane component of angular momentum, $m$,
$h.c.$ stands for the Hermitian conjugation, and $\ldots$ denotes the remaining 4 terms obtained from the general expression  Eq. \eqref{eq:3rd}.
All such terms produce a similar contribution.
In result we estimate,
\begin{align}
    \lambda \approx g_L \left(\frac{\lambda_{\text{SO}}   }{\Delta E_{cr}} \right)
    \left( \frac{\Delta E_{str}}{\Delta E_{cr}} \right)\, , 
\end{align}
where $\Delta E_{cr}$ is the typical spin splitting and $\Delta E_{str}$ is the typical energy scale associated with the strain perturbation $\mathcal{H}_{\varepsilon}$.

\section{Critical field of a superconductor with an anisotropic $g$-factor}
\label{app:Bc}

Here we study the effect of the anisotropy of the $g$-factor as expressed by Eq.~\eqref{eq:H_zr} on the critical field of a superconductor.
In particular we derive the expressions \eqref{eq:beta_anis} and \eqref{eq:alpha_anis} controlling the critical field and the two-fold anisotropy.  
We assume the superconductor is described by the Hamiltonian Eq.~\eqref{eq:H0}.

The $4\times4$ Green function $\hat{G}\left(\mathbf{k};\omega_{n}\right)$ satisfies the Gor'kov equation,
\begin{equation}
\left[i\omega_{n}\hat{\sigma}_{0}-\hat{H}_{\mathrm{BdG}}-\hat{\Sigma}\right]\hat{G}\left(\mathbf{k};\omega_{n}\right)=\hat{\sigma}_{0}\label{eq:B.2}
\end{equation}
where $\sigma_{0}$ is the $2\times2$ unit matrix,  $\omega_{n}=\pi T\left(2n+1\right)$ are the Matsubara frequencies and $\hat{H}_{\mathrm{BdG}}$ is the Bogoliubov--de-Gennes (BdG) Hamiltonian corresponding to the normal state Hamiltonian [Eq.~\eqref{eq:H0}],
\begin{equation}\label{eq:BdG}
\hat{H}_{\mathrm{BdG}} \!=\!\left[\begin{array}{cc}
\xi_{\mathbf{k}}+\left[\boldsymbol{\gamma}\left(\mathbf{k}\right)-\mathbf{B}_{\mathrm{eff}}\right]\!\cdot\!\boldsymbol{\sigma} & \Delta\\
\Delta^{\dagger} & -\xi_{\mathbf{k}}+\left[\boldsymbol{\gamma}\left(\mathbf{k}\right)+\mathbf{B}_{\mathrm{eff}}\right]\!\cdot\!\boldsymbol{\sigma}^{\mathrm{T}}
\end{array}\right]\, ,
\end{equation}
where the Zeeman field $\mathbf{B}$ is replaced by $\mathbf{B}_{\mathrm{eff}}$ introduced in Eq.~\eqref{eq:H_zr} to incorporate the effect of strain.

As in the main text, in Eq.~\eqref{eq:BdG} the isotropic OP coincides with the spectral gap in the BCS limit, $\Delta=\psi i\sigma_{2}$.
The self-energy $\hat{\Sigma}$ is due to the disorder scattering, 
\begin{equation}\label{eq:Sigma}
\hat{\Sigma}=\Gamma\int\frac{\mathrm{d}\varphi_{\mathbf{k}}}{2\pi}\int\frac{d\xi_{\mathbf{k}}}{\pi}\hat{\sigma}_{z}\hat{G}\left(\mathbf{k};\omega_{n}\right)\hat{\sigma}_{z}
\end{equation}
where $\Gamma$ is the scattering rate off the scalar disorder, $\hat{\sigma}_{z}=\mathrm{diag}\left(\sigma_{0},-\sigma_{0}\right)$, and
$\tan \varphi_{\mathbf{k}} = k_y/k_x$.
We introduce the quasi-classical Green function in the form,
\begin{align}\label{eq:g}
\hat{g}\left(\mathbf{k}_{\mathrm{F}}\right) & =\int_{-\infty}^{\infty}\frac{\mathrm{d}\xi_{\mathbf{k}}}{\pi}i\hat{\sigma}_{z}\hat{G}\left(\mathbf{k};\omega_{n}\right)\\
 & =\left[\begin{array}{cc}
g\left(\mathbf{k}_{\mathrm{F}};\omega_{n}\right) & -if\left(\mathbf{k}_{\mathrm{F}};\omega_{n}\right)\\
-if^{*}\left(-\mathbf{k}_{\mathrm{F}};\omega_{n}\right) & -g^{*}\left(-\mathbf{k}_{\mathrm{F}};\omega_{n}\right)
\end{array}\right],\notag
\end{align}
where $\mathbf{k}_{\mathrm{F}} = \bm{k}/k$.
We parametrize the function, $f\left(\mathbf{k}_{\mathrm{F}};\omega_{n}\right)$ in Eq.~\eqref{eq:g} in the standard form as follows \cite{Eschrig2015},
\begin{equation}\label{eq:f_param}
f\left(\mathbf{k}_{\mathrm{F}};\omega_{n}\right)=\left[f_{0}\left(\mathbf{k}_{\mathrm{F}};\omega_{n}\right)\sigma_{0}+\mathbf{f}\left(\mathbf{k}_{\mathrm{F}};\omega_{n}\right)\cdot\boldsymbol{\sigma}\right]i\sigma_{2}\, .
\end{equation}
To find the critical field, it is sufficient to evaluate the functions, $f_{0}$ and $\mathbf{f}$ to the linear order in $\psi$ denoted here by $f_{0}^{\left(1\right)}$ and $\mathbf{f}^{\left(1\right)}$, respectively.
These expressions can be found from the Eilenberger equation lineraized in the OP in the form \cite{Haim2020}
\begin{subequations}\label{eq:Eilen_lin}
\begin{align}\label{eq:f0_1}
\omega_{n}f_{0}^{\left(1\right)}= &
i\mathbf{f}^{\left(1\right)}\cdot\mathbf{B}_{\mathrm{eff}}+\mathrm{sgn}\left(\omega_{n}\right)\psi
\notag \\
& +\Gamma\mathrm{sgn}\left(\omega_{n}\right)\left[\left\langle f_{0}^{\left(1\right)}\right\rangle -f_{0}^{\left(1\right)}\right],
\end{align}
\begin{align}\label{eq:f_1}
\omega_{n}\mathbf{f}^{\left(1\right)}=
&
if_{0}^{\left(1\right)}\mathbf{B}_{\mathrm{eff}}+\boldsymbol{\gamma}_{\mathbf{k}}\times\mathbf{f}^{\left(1\right)}
\notag \\
& +\Gamma\mathrm{sgn}\left(\omega_{n}\right)\left[\left\langle \mathbf{f}^{\left(1\right)}\right\rangle -\mathbf{f}^{\left(1\right)}\right].
\end{align}
\end{subequations}
where $\left\langle \cdots\right\rangle = \int d \varphi_{\mathbf{k}}/ 2 \pi$ stands for the angular average over the Fermi surface. 

The critical field is determined by the self-consistency equation written to the first order in the OP \cite{Mockli2020}
\begin{align}\label{eq:self-cons1}
\ln\left(\frac{T}{T_{c}}\right)+\pi T\sum_{n=-\infty}^{\infty}\left(\frac{1}{\left|\omega_{n}\right|}-\frac{1}{\psi}\left\langle f_{0}^{\left(1\right)}\right\rangle \right)=0. 
\end{align}
To simplify the calculations we consider the SO coupling of the form,
$\boldsymbol{\gamma}\left(\varphi_{\mathbf{k}}\right)=\hat{z} \gamma^I \mathrm{sgn\left[\cos\left(3\varphi_{\mathbf{k}}\right)\right]}$.
Such a coupling is appropriate for the SO coupling at $K$ pockets.
Our results are qualitatively unchanged with other $\bm{k}$ dependence of the Ising SO coupling.
With the above choice of the SO coupling the solution to Eq.~\eqref{eq:Eilen_lin}
reads
\begin{equation}
\label{eq:f0_av}
\left\langle f_{0}^{(1)}\right\rangle =\frac{\psi\left(\left(\gamma^{I}\right)^{2}+\Gamma\left|\omega_{n}\right|+\omega_{n}^{2}\right)}{\left|\omega_{n}\right|\left(B_{\mathrm{eff}}^{2}+\left(\gamma^{I}\right)^{2}+\omega_{n}^{2}\right)+\Gamma\left(B_{\mathrm{eff}}^{2}+\omega_{n}^{2}\right)}. 
\end{equation}
Substituting Eq.~\eqref{eq:f0_av} into Eq.~\eqref{eq:self-cons1} gives an equation \eqref{eq:self-cons_Bc} of the main text.
The second order expansion of Eq.~\eqref{eq:f0_av} in $B_{\mathrm{eff}}$ yields Eq.~\eqref{eq:beta_anis} for the coefficient $\beta$.

\end{appendix}


%

\end{document}